\documentclass{aa}
\usepackage[varg]{txfonts}
\usepackage{natbib}
\bibpunct{(}{)}{;}{a}{}{,}

\usepackage[colorlinks, citecolor=blue, linkcolor=blue, urlcolor=black]{hyperref}
\usepackage{graphicx}
\usepackage{xcolor}
\usepackage{siunitx}

\usepackage{chemformula}
\RenewChemArrow{->}{\hspace{0.225em}$\longrightarrow$}
\setchemformula{arrow-offset=0.7em,compound-sep=0.225em}

\usepackage{hypcap}
\usepackage{placeins}
\usepackage{orcidlink}

\begin{document}

\title{A sensitivity analysis of interstellar ice chemistry in astrochemical models}

\author{Tobias~M.~Dijkhuis\inst{\ref{LIC},\ref{LO},\ref{IMM}}\fnmsep\thanks{Corresponding authors: \url{t.m.dijkhuis@lic.leidenuniv.nl}; \url{a.l.m.lamberts@lic.leidenuniv.nl}}\orcidlink{0009-0009-2498-6429}
\and Thanja~Lamberts\inst{\ref{LIC},\ref{LO}}\fnmsep\footnote[1]{}\orcidlink{0000-0001-6705-2022}
\and Serena~Viti\inst{\ref{LO},\ref{Bonn},\ref{UCL}}\orcidlink{0000-0001-8504-8844}
\and Herma~M.~Cuppen\inst{\ref{IMM}}\orcidlink{0000-0003-4397-0739}}

\institute{Leiden Institute of Chemistry, Gorlaeus Laboratories, Leiden University, PO Box 9502, 2300 RA
Leiden, The Netherlands\label{LIC} \and
Leiden Observatory, Leiden University, PO Box 9513, 2300 RA Leiden, The Netherlands \label{LO} \and
Institute for Molecules and Materials, Radboud University, 6525 AJ Nijmegen, The Netherlands \label{IMM} \and
Transdisciplinary Research Area (TRA) `Matter'/Argelander-Institut für Astronomie, University of Bonn, 53121 Bonn, Germany \label{Bonn} \and
Department of Physics and Astronomy, University College London, Gower Street, London, UK \label{UCL}
}

\date{Received date /
Accepted date }

\abstract
{
Astrochemical models are essential to bridge the gap between the timescales of reactions, experiments, and observations. Ice chemistry in these models experiences a large computational complexity as a result of the many parameters required for the modeling of chemistry occurring on these ices, such as binding energies and reaction energy barriers. Many of these parameters are poorly constrained, and accurately determining all would be too costly.
}
{
We aim to find out which parameters describing ice chemistry have a large effect on the calculated abundances of ices for different prestellar objects.
}
{
Using Monte Carlo sampled binding energies, diffusion barriers, desorption and diffusion prefactors, and reaction energy barriers, we determined the sensitivity of the abundances of the main ice species calculated with UCLCHEM, an astrochemical modeling code, on each of these parameters. We do this for a large grid of physical conditions across temperature, density, cosmic ray ionization rate and UV field strength.
}
{
Regardless of the physical conditions, the main sensitivities of abundances of the main ice species are the diffusion barriers of small and relatively mobile reactive species such as \ch{H}, \ch{N}, \ch{O}, \ch{HCO}, and \ch{CH3}. Thus, these parameters should be determined more accurately to increase the accuracy of models, paving the way to a better understanding of observations of ices. In many cases, accurate reaction energy barriers are not essential due to the treatment of competition between reactions and diffusion.
}
{}


\titlerunning{A sensitivity analysis of interstellar ice chemistry}
\authorrunning{T.~M.~Dijkhuis et al.}

\maketitle

\section{Introduction}
\nolinenumbers
\label{sec:Introduction}
Astrochemical modeling is essential to bridge the gap between the time scales of chemical reactions occurring on the order of $10^{-9}$~seconds, experiments taking minutes to days, and the evolution of astronomical objects over about $10^6$~years. It aids the interpretation of the 
snapshot information obtained from observations, both by being able to determine the physical conditions such as the temperature and density, and by simulating the time evolution of the observed object.

Interstellar ices play a crucial role in the chemical evolution of dark molecular clouds and prestellar cores, as they allow reactions to take place that cannot take place in the gas phase. Namely, icy grains provide a third body to dissipate excess energy released by reactions, and simultaneously act as a reactant concentrator. The chemistry is not only influenced by depleting atoms and molecules from the gas phase, but also contribute more chemical diversity by sublimating molecules that formed on the ice back to the gas. Without icy grains, even the simplest molecule, \ch{H2}, would not exist in the observed abundance \citep{Gould1963}. Chemistry on these ices is governed by various competing processes, such as adsorption (species colliding with the grain and sticking to it), desorption (sublimation from the ice to the gas-phase), diffusion (or hopping, species moving around on and in the ice), and reactions (breaking and formation of chemical bonds). Many molecules, such as \ch{H2O}, \ch{CO2} and \ch{CH3OH}, are formed mostly on the ices by a combination of these processes.

However, the parameters necessary to describe ice chemistry in astrochemical models are poorly constrained. Modeling ice chemistry, at minimum, requires the binding energies of all species to the ice surface and the energy barriers of all reactions. While there are databases available for gas-phase reactions and their rate constants, the most commonly used of which are UDfA \citep{Millar2024} and KIDA \citep{Wakelam2024}, many binding energies and reaction energy bariers are unknown. Because of computational and experimental limitations, these are often harder to obtain than gas-phase reaction rate constants. Thus, historically, many have been fitted from observations or roughly estimated based on chemical intuition. For example, reactions between two radicals are often assumed to be barrierless, however recent works have shown that this is not necessarily the case, as some geometries exhibit large barriers \citep{EnriqueRomero2022,Molpeceres2023} or lead to no reaction at all \citep[see for example][]{Lamberts2019}. Another example is the estimation of diffusion barriers from binding energies using a universal ratio of $\chi=E_{\mathrm{diff}}/E_{\mathrm{bind}}$, because of a lack of accurate diffusion barriers for many species \citep{Furuya2022, Ligterink2025}.

There is clearly a need to obtain the grain-chemistry parameters more accurately. While this is a daunting task given the large amount of parameters required in a model, fortunately not all parameters are equally important to constrain. A sensitivity analysis can aid in identifying which parameters are crucial for accurate models. One of the first examples of such an analysis for astrochemical models is the work of \citet{Kuntz1976}, who investigated the effect of varying gas-phase reaction rate constants on steady-state abundances in the interstellar medium. Since then, many sensitivity analyses focusing on the gas-phase reaction rate constants have been performed for various types of astronomical objects \citep[see for example][]{Wakelam2006,Vasyunin2008, Wakelam2010,Byrne2024}.

Some sensitivity analyses of ice chemistry have also been performed. \citet{Penteado2017} varied the binding energies of many ice species to gain information about the effect that the binding energies have on main ice abundances, like \ch{H2O}, \ch{CO2}, and \ch{CH3OH}, in a typical dense molecular cloud. Because of the fixed $E_{\mathrm{diff}}/E_{\mathrm{bind}}$ ratio, diffusion barriers were indirectly altered as well. \citet{Iqbal2018} performed a similar study, again for a dark molecular cloud, where they varied the diffusion barriers of all species. Finally, \citet{Furuya2022} also varied the diffusion barriers of ice species, but used more realistic time-dependent density and temperature profiles. Sensitivity analyses can also be performed using Bayesian methods based on Markov chain Monte Carlo algorithms \citep{Holdship2018} or coupled with machine learned statistical emulators \citep{Heyl2022,Vermarien2025}.

This work aims to obtain sensitivities of the most abundant ice species not only on binding energies, but also on diffusion barriers, reaction energy barriers, along with diffusion and desorption prefactors, to gain information on exactly which chemical parameters should be further constrained with computational or experimental chemical tools. We do this for a large grid of varying physical conditions corresponding to different astronomical objects, ranging from early stages of a molecular cloud to prestellar cores.

The structure of this paper is the following: Sect.~\ref{sec:chemical_model} introduces UCLCHEM, the chemical model used in this work, and discusses the changes made to it to treat ice chemistry more accurately by implementing recent insights. Then, Sect.~\ref{sec:sensitivity_analysis} discusses the methodology for the sensitivity analysis and the physical conditions studied here. In Sect.~\ref{sec:Results} we show and discuss the results, Sect.~\ref{sec:implications} discusses implications for all three pillars of astrochemistry, and finally in Sect.~\ref{sec:summary_and_conclusions} we summarize the conclusions.

\section{Chemical model}
\label{sec:chemical_model}
UCLCHEM\footnote{\url{https://github.com/uclchem/UCLCHEM}} is a fully open-source time-dependent gas-grain chemical code \citep{Holdship2017}. It uses the three-phase, meaning gas, surface ice and bulk ice, formulation of \citet{Garrod2011} and uses \citet{Garrod2013} to take into account the swapping between surface and bulk. The gas and dust temperature are assumed to be the same. We use the gas-phase reactions from UDfA Rate22 \citep{Millar2024}. The reaction rates are integrated using DVODE \citep{Brown1989}. 

Because we aim to study the sensitivity of the abundances to chemical parameters at various physical conditions, a constant temperature and density is used throughout individual simulations. A list of all species and ice reactions in the network can be found on the GitHub repository\footnote{\url{https://github.com/uclchem/SensitivityAnalysisIces}}. Table~\ref{tab:elementalAbundances} lists the elemental abundances used here. All elements are initially in atomic form, except for hydrogen, of which 50\% is in \ch{H2}, and carbon, which is all singly ionized, \ch{C+}/\ion{C}{ii}. We use spherical grains with a radius of 0.1~$\muup$m, a density of 3~g~cm$^{-3}$, and a surface site density of $1.5\times10^{15}$~cm$^{-2}$. The upper two layers of ice are treated as the surface, and the rest is considered bulk. We assume that molecular hydrogen is not pushed to the bulk by the accretion of new material because of its rapid desorption rate. Species in the bulk of the ice have the same binding energy and diffusion barrier as water to ensure that diffusion of bulk species is limited by self-diffusion of water, as found by \citet{Ghesquire2015}. The gas-to-dust mass ratio is taken as 100 \citep{Kimura2003}. Both thermal and nonthermal desorption mechanisms are included, and the latter are described in \citet{Holdship2017}. Cosmic ray attenuation \citep[see e.g.,][]{Padovani2018} is switched off because we want to determine the effect of cosmic rays on the chemistry regardless of the density.
\begin{table}
    \caption{Elemental abundances used in this work.}
    \label{tab:elementalAbundances}
    \centering
    \sisetup{table-format = 1.3e1, table-alignment-mode = format}
    \begin{tabular}{cS[table-number-alignment=center]}
    \hline\hline
        Element & \multicolumn{1}{c}{Abundance (with respect to H)} \\
        \hline
         \ch{He} & \multicolumn{1}{c}{0.5} \\
         \ch{C} & 1.77e-4 \\
         \ch{N} & 6.18e-5 \\
         \ch{O} & 3.34e-4 \\
         \ch{F} & 3.6e-8 \\
         \ch{Mg} & 2.256e-6 \\
         \ch{Si} & 1.78e-6 \\
         \ch{P} & 7.78e-8 \\
         \ch{S} & 3.51e-6  \\
         \ch{Cl} & 3.39e-8 \\
         \ch{Fe} & 2.01e-7 \\
         \hline
    \end{tabular}
    \tablefoot{Abundances taken from the heavily depleted case ($F_{*}=1$) in Table~4 of \citet{Jenkins2009}. Abundance of F taken from Table~1 of \citet{Asplund2009} in present-day solar photosphere.}
\end{table}

The rate coefficient for thermal desorption (\ch{\textit{i}\textsuperscript{ice} -> \textit{i}\textsuperscript{gas}}) is calculated as
\begin{equation}
    k_{\text{des}}^{i} = \nu_{\text{des}}^{i}\exp\left(-\frac{E_{\text{bind}}^{i}}{T_{\text{dust}}}\right),
\end{equation}
where $\nu_{\text{des}}^{i}$ is the prefactor for desorption of species $i$, and $E_{\text{bind}}^i$ is its binding energy in K. The desorption prefactor is usually calculated as \citep{Hasegawa1992,Tielens1987}
\begin{equation}
    \nu_{\text{des}}^i = \sqrt{\frac{2n_sk_{\mathrm{B}}E_{\text{bind}}^i}{\pi^2m^i}},
    \label{eq:desprefacHH}
\end{equation}
where $n_s$ is the surface site density on the grains, and $m^i$ is the mass of $i$. This equation results in values of around $10^{12}$\textendash$10^{13}$~s$^{-1}$ for most species. However, this neglects the contribution of the rotational partition function, which can lead to underestimation of the prefactor for molecular species by multiple orders of magnitude \citep[see for example][]{Minissale2022,Ligterink2023}.

The diffusion rate coefficient of species $i$ on the surface is similarly calculated as
\begin{equation}
    k_{\text{diff}}^{i} = \nu_{\text{diff}}^{i}\exp\left(-\frac{E_{\text{diff}}^{i}}{T_{\text{dust}}}\right),
    \label{eq:hoppingRate}
\end{equation}
where $\nu_{\text{diff}}^{i}$ and $E_{\text{diff}}^i$ are the prefactor and energy barrier for diffusion of species $i$, respectively. The prefactor for diffusion is usually assumed to be the same as the desorption prefactor. Whereas the diffusion barrier is often assumed to be a fixed fraction of the binding energy, we have decoupled the binding energy and diffusion barrier in UCLCHEM such that they can be varied independently, and more accurate diffusion barriers can be supplied to the model in the future. Equation~\ref{eq:hoppingRate} assumes diffusion is a completely thermal process, that is to say, tunneling is not relevant for diffusion. \citet{Asgeirsson2017} and \citet{Senevirathne2017} showed that this is a valid approximation for diffusion of atomic hydrogen on amorphous solid water at temperatures down to as low as 10~K. Thus, even more-so for heavier species, tunneling on amorphous solid water will be completely negligible. Considering that we are looking at temperatures far below the crystallization temperature of \ch{H2O} of about 150~K \citep{Lofgren2002,May2012}, this approximation will be used here.

Reactions on the ice mostly occur through the Langmuir-Hinshelwood mechanism, where two species diffuse on the grain surface, meet each other and can react \citep{Cuppen2017}. The rate constant for a Langmuir-Hinshelwood reaction between species $i$ and $j$, \ch{\textit{i}\textsuperscript{ice} + \textit{j}\textsuperscript{ice} -> \textit{k}\textsuperscript{ice}}, is determined using the formalism introduced by \citet{Hasegawa1992}, as
\begin{equation}
    \kappa^{i+j} = \max\left[\nu_{\text{diff}}^i, \nu_{\text{diff}}^{j}\right]P_{\text{reac}}^{i+j},
    \label{eq:reacRateNoComp}
\end{equation}
where $P_{\text{reac}}^{i + j}$ is the reaction probability upon species $i$ and $j$ meeting, calculated by either a Boltzmann factor or the tunneling probability through a rectangular and symmetric barrier,
\begin{equation}
    P_{\text{reac}}^{i + j} = \max\left[\exp\left(-\frac{E_{\text{reac}}^{i + j}}{T_{\text{dust}}}\right),\exp\left(-\frac{2a}{\hbar}\sqrt{2\mu^{i + j} k_{\mathrm{B}}E_{\text{reac}}^{i + j}}\right)\right],
    \label{eq:reactionProbability}
\end{equation}
where $a$ is the barrier width, assumed to be 1.4~\AA{} for all reactions, and $\mu^{i + j}$ and $E_{\text{reac}}^{i + j}$ are the mass of the tunneling particle and energy barrier for the reaction between species $i$ and $j$, respectively. The treatment of tunneling and the value of $\mu^{i + j}$ are further discussed in App.~\ref{app:tunneling_mass}. A reaction is referred to as barrierless if $E_{\mathrm{reac}}^{i + j}=0$~K, such that $P_{\mathrm{reac}}^{i+j}=1$, which means that the reaction has a 100\% chance of occurring when the two reagents meet. The rectangular barrier approximation results in a temperature-independent tunneling probability, which is not always the case \citep[see for example Fig.~7 in][]{Lamberts2017}.

The competition between reaction, diffusion, and thermal desorption is taken into account by multiplying Eq.~\ref{eq:reacRateNoComp} with the probability that the reaction actually occurs when the two reagents meet, and either species does not diffuse away or desorb instead \citep{Chang2007},
\begin{equation}
    f_{\text{comp}} = \frac{\kappa^{i + j}}{\kappa^{i + j} + k_{\text{diff}}^i + k_{\text{diff}}^{j} + k_{\text{des}}^i + k_{\text{des}}^j}.
    \label{eq:competitionFraction}
\end{equation}
In practice, $f_{\text{comp}}$ is close to unity if the reaction is faster than the hopping timescale of the reactants, that is to say, if $\kappa^{i+j}>k_{\mathrm{diff}}^{i/j}$. The final reaction rate constant is then
\begin{equation}
    k_{\text{reac}}^{i + j} = \alpha^{i+j\rightarrow k}\frac{f_{\text{comp}}\left(k_{\mathrm{diff}}^{i}+k_{\mathrm{diff}}^{j}\right)}{N_{\mathrm{sites}}X_{\text{dust}}},
    \label{eq:rateConstantInclCompetition}
\end{equation}
where $N_{\mathrm{sites}}$ is the number of sites per dust grain and $X_{\text{dust}}$ is the abundance of dust grains with respect to hydrogen nuclei. $\alpha^{i+j\rightarrow k}$ is the branching ratio, which is not unity if, for example, a reaction between $i$ and $j$ can form either $k$ or $l$. Rate constants of reactions in bulk ice are scaled by $1/N_{\text{ML}}^{\text{bulk}}$, or in other words, are divided by the amount of monolayers of bulk ice, to take into account that it will take a particle $N_{\text{ML}}^{\text{bulk}}$ times longer to scan the entire bulk ice \citep{Ruaud2016}. The rate of change in the fractional abundance of $i$ in units of fractional abundance s$^{-1}$ as a result of the reaction between $i$ and $j$ forming $k$ is then
\begin{equation}
    \left[\frac{\mathrm{d}X^i}{\mathrm{d}t}\right]_{i+j}=-k_{\text{reac}}^{i+j}X^iX^j,
\end{equation}
where $X^i$ and $X^j$ are the abundances of $i$ and $j$ (differentiated between surface and bulk ice) with respect to hydrogen nuclei.

\subsection{Chemical desorption}
\label{subsec:ChemicalDesorption}
Exothermic reactions are assumed to lead to the desorption of one of the products by transferring some of the released energy to the translation away from the surface (\ch{$i^{\text{ice}}$ + $j^{\text{ice}}$ -> $k^{\text{gas}}$}). This process is called ``chemical'' desorption (CD), also sometimes referred to as reactive desorption. Here, we use a combination of the approximations introduced by \citet{Minissale2015} for bare grains and \citet{Fredon2021} for ices. The treatment of \citet{Minissale2015} assumes that when the reaction has occurred, the resulting products first collide elastically with the surface, and bounce back to then desorb. In the case of a two-product reaction, the released energy is distributed between the products $k$ and $l$ according to their masses, similarly to \citet{Riedel2023},
\begin{equation}
    w^k = \frac{m^l}{m^k+m^l},
    \label{eq:distributionEnergyFromMasses}
\end{equation}
where $m$ indicates the mass of the product. $w^k$ is unity for a one-product reaction. The fraction of energy released from the reaction that product $k$ contains after its collision with the bare grain is
\begin{equation}
    \epsilon_{\text{grain}}^k = \left(\frac{M - m^k}{M + m^k}\right)^2w^k,
\end{equation}
where $M$ is an effective mass of the grain, which \citet{Minissale2015} found to be about 120~amu to match experiments. Then, the probability that species $k$ desorbs from a bare grain is 
\begin{equation}
    \eta_{\text{CD,grain}}^k = \exp\left(-\frac{3N_{\text{atoms}}^kE_{\text{bind}}^k}{\epsilon_{\text{grain}}^k \Delta H_\text{reac}}\right),
    \label{eq:CDefficiency}
\end{equation}
where $E_{\text{bind}}^k$ is its binding energy and $N_{\text{atoms}}^k$ is the number of atoms in $k$. $\Delta H_\text{reac}$ is the enthalpy of reaction with opposite sign, that is, the amount of energy released from the reaction,
\begin{equation}
    \Delta H_{\text{reac}}= \Delta H_{\text{form}}^i + \Delta H_{\text{form}}^j - \Delta H_{\text{form}}^k
\end{equation}
for a one-product reaction. Enthalpies of formation $\Delta H_{\text{form}}$ were taken from KIDA \citep{Wakelam2024}, CCCBDB \citep{CCCBDB}, or CDMS \citep{Muller2005}, or if unavailable were calculated as described in App.~\ref{app:formation}. If $\epsilon_{\text{grain}}^k\Delta H_{\mathrm{reac}}$ is smaller than the binding energy of species $k$, chemical desorption is not possible, meaning $\eta_{\mathrm{CD,grain}}^{k}=0$. While the same equation, Eq.~\ref{eq:CDefficiency} is different from what was previously implemented in UCLCHEM \citep[see][]{Quenard2018}, because the previous implementation assumed that in the case of a two-product reaction, either both products desorb or both stay on the grain, and included the energy barrier of the reaction in the calculation of $\Delta H_{\text{reac}}$.

While this treatment might be accurate on a bare grain, it fails to capture chemical desorption on ices because it assumes an elastic collision with the grain surface, as shown by \citet{Fredon2017}. Thus, for the chemical desorption probability from ices we use the treatment from \citet{Fredon2021}, where
\begin{equation}
    \eta_{\text{CD,ice}}^k = f\left(1-\exp\left[-\frac{\epsilon_{\text{ice}}^k\Delta H_{\text{reac}}-E_{\text{bind}}^k}{3E_{\text{bind}}^k}\right]\right).
\end{equation}
Here, $\epsilon_{\text{ice}}^k\Delta H_{\text{reac}}$ is the amount of energy transferred to the translational degrees of freedom of product $k$, where $\epsilon_{\text{ice}}^k$ is
\begin{equation}
    \begin{split}
    \epsilon_{\text{ice}}^k = \left\{
        \begin{array}{l l}
            \epsilon_{\text{ice},1} & \text{ for one-product reactions, and} \\
            \epsilon_{\text{ice},2}\cdot w^k & \text{ for two-product reactions.}\\
        \end{array}\right.
    \end{split}
\end{equation}
\citet{Furuya2022ChemDes} found that $\epsilon_{\text{ice},1}=0.07$ best fitted experimental chemical desorption data, so that is what we use here. This also accurately captures several theoretical energy dissipation studies. For example, for \ch{H + CO -> HCO}, $\epsilon_{\text{ice},1} \approx 0.073$\textendash$0.11$ \citep{Pantaleone2020}, and for \ch{H + P -> PH} $\epsilon_{\text{ice},1}\approx 0.01$\textendash$0.05$ \citep{Molpeceres2023}. This is still a strongly simplified approach, and is not accurate for all reactions. For example, for \ch{H + N -> NH}, $\epsilon_{\text{ice},1}$ was found to be about 0.007 because of its very efficient energy dissipation to the ice \citep{Ferrero2023}. As recommended by \citet{Fredon2021}, we use $\epsilon_{\text{ice},2}=0.2$ and $f=0.5$. If $\epsilon_{\text{ice}}^k\Delta H_{\text{reac}} < E_{\text{bind}}^k$, no chemical desorption from ices is allowed, meaning $\eta_{\text{CD,ice}}^k=0$, according to \citet{Fredon2017}. 

The overall chemical desorption efficiency is calculated by linearly interpolating between the grain efficiency and ice efficiency as
\begin{equation}
    \eta_{\text{CD}}^k = \eta_{\text{CD,grain}}^k + \min\left[N_{\text{ML}}^{\text{ice}}, 1.0\right]\left(\eta_{\text{CD,ice}}^k-\eta_{\text{CD,grain}}^k\right),
\end{equation}
where $N_{\text{ML}}^{\text{ice}}$ is the number of monolayers of total ice on the dust grain. The rate constant of the reaction between $i$ and $j$ where product $k$ desorbs is then
\begin{equation}
    k_{\text{CD}}^k = \eta_{\text{CD}}^k k_{\text{reac}}^{i + j}.
\end{equation}
The rate constant of the same reaction where all products stay on the grain is then calculated as
\begin{equation}
    k_{\text{no CD}}^{i + j} = k_{\text{reac}}^{i + j}\left(1-\textstyle\sum_{k \in \text{products}}\eta_{\text{CD}}^k\right).
\end{equation}
In the case of a reaction with multiple products, if $\sum_{k \in \text{products}}\eta_{\text{CD}}^k>1$, then $k_{\text{no CD}}^{i + j}<0$. This corrects the formation of species $k$ on the grain by the reaction where species $l$ chemically desorbs, and vice versa. Because of the more rapid energy dissipation and higher binding energies, chemical desorption from a reaction in bulk ice is not possible.

\subsection{Encounter desorption}
\label{subsec:encounter_desorption}
At relatively high densities of $n_{\mathrm{H}}\geq 10^5$~cm$^{-3}$ and 10~K, models overpredict the build-up of molecular hydrogen in the ice, because of the amount of hydrogen in the gas-phase. However, this is not physically accurate, because the \ch{H2} on \ch{H2} binding energy \citep[23~K,][]{Cuppen2007} is too low for a multilayer ice of \ch{H2} to form, as discussed for example by \citet{Morata2013}. This artificial amount of \ch{H2} ``ice'' results in a different and incorrect ice chemistry, as discussed by, for example \citet{Penteado2017}. To remedy this, the ``encounter'' desorption (ED) mechanism was proposed by \citet{Hincelin2015}. When one hydrogen molecule diffuses over another, its binding energy sharply decreases to 23~K, strongly increasing its desorption probability. The rate constant for this reaction is calculated as 
\begin{equation}
    k_{\text{ED}} = \frac{2k_{\text{diff}}^{\text{H}_2}}{N_{\text{sites}}X_{\text{dust}}}\frac{k_{\text{des}}^{\text{H}_2\text{ on H}_2}}{k_{\text{des}}^{\text{H}_2\text{ on H}_2} + k_{\text{diff}}^{\text{H}_2\text{ on H}_2}},
\end{equation}
where the second term captures the competition between desorption and diffusion, and the superscript ``\ch{H2} on \ch{H2}'' indicates the use of the binding energy of \ch{H2} on a \ch{H2} surface of 23~K, and diffusion barrier of $0.5\times 23=11.5$~K. The resulting rate of change of the surface abundance of \ch{H2} is then calculated similarly as that of a Langmuir-Hinshelwood reaction between two \ch{H2} molecules, \ch{H2\textsuperscript{surf} + H2\textsuperscript{surf} -> H2\textsuperscript{gas} + H2\textsuperscript{surf}},
\begin{equation}        
    \left[\frac{\mathrm{d}X^{\text{H}_2}}{\mathrm{d}t}\right]_{\text{ED}}=-\frac{1}{2}k_{\text{ED}}X^{\text{H}_2}X^{\text{H}_2}.
\end{equation}
We chose not to use the method from \citet{Garrod2011} because it makes the ordinary differential equations stiffer, increasing computational cost.

\section{Sensitivity analysis}
\label{sec:sensitivity_analysis}
\subsection{Physical grid}
\label{subsec:physical_grid}

Sensitivities of the chemistry to different chemical parameters are obtained at different physical conditions using a 4-dimensional grid with changing temperature $T$, hydrogen number density $n_{\text{H}}$, cosmic ray ionization rate $\zeta$, and UV field strength $F_{\mathrm{UV}}$. While $\zeta$ and $F_{\mathrm{UV}}$ do not necessarily depend on the type of astronomical object, they can have a large effect on the regime that the chemical network is pushed into, and thus using a low or high value might result in different sensitivities. The values used for this 4-dimensional grid can be found in Table~\ref{tab:physicalGrid}, resulting in $5\times4\times3\times3\ +\ 5\times3\times1\times3 = 225$ total grid points. 
\begin{table}
\caption{Grid of physical conditions studied.}
\label{tab:physicalGrid}
\centering
\begin{tabular}{c l l}
    \hline\hline
    Parameter & Values & Unit \\
    \hline
    $T$ & 10, 20, 30, 40, 50 & K \\
    $n_{\text{H}}$ & $10^{3}$, $10^{4}$, $10^5$, $10^6$ & cm$^{-3}$ \\
    $\zeta$\tablefootmark{a} & $10^{-1}$, $10^0$, $10^1$, $10^2$ & $1.3\times10^{-17}$ s$^{-1}$ \\
    $F_{\mathrm{UV}}$ & $10^{-1}$, $10^0$, $10^1$ & Habing \\
    \hline
\end{tabular}
\tablefoot{
    \tablefoottext{a}{$\zeta=1.3\times10^{-15}$~s$^{-1}$ was not run at $n_{\mathrm{H}}=10^6$~cm$^{-3}$, see main text.}
}
\end{table}
The ``size'' of the object, used to calculate the column density and thus the visual extinction, is kept the same across model runs, such that models with a higher density have a higher visual extinction. We attempted to run the grid at the highest density and cosmic ray ionization rate, but because of the rapid freeze-out and nonthermal desorption at these extreme conditions, the ordinary differential equations became extremely stiff and some models did not finish. Regardless, we expect these conditions to be rare in real astronomical environments, because the high cosmic ray ionization rate would heat and ionize the gas to prevent the collapse to higher densities.

\subsection{Sampling chemical parameters}
\label{subsec:sampling_chemical_parameters}
For each point in our physical grid, a sensitivity analysis of the calculated abundances is performed on the binding energies, diffusion barriers, and diffusion and desorption prefactors of each ice species and reaction energy barriers of each Langmuir-Hinshelwood reaction in the network. This results in $d=118\times4+158=630$ chemical parameters. We chose to vary the energies and prefactors (rather than vary the rate constants $k$ directly) because they are the input in astrochemical models. Because chemical reaction networks are a set of nonlinear ordinary differential equations \citep[see][]{Angeli2009}, a one-at-a-time sampling method, where a single parameter is changed from its default value while the others are kept fixed, is not sufficient \citep{Saltelli2019}. Moreover, because of the high dimensionality, it would require many model samples. 

Instead, a global Monte Carlo approach is used, where all parameters are varied simultaneously. Because this is a global method, couplings between different parameters are correctly recovered \citep[see for example][]{Dobrijevic2010}. First, 1000 uniformly distributed random points are generated in the $d$-dimensional unit hypercube, $\left[0,1\right)^d$. Each point $\vec{x}_i$ is then transformed such that the distribution of all points corresponds to a truncated, between $\vec{y}^{\text{min}}$ and $\vec{y}^{\text{max}}$, normal distribution around mean $\vec{\mu}$ and with standard deviation $\vec{\sigma}$ by the inverse transform method:
\begin{align}
    \vec{p}_i &= \left[\Phi\left(\vec{y}^{\text{max}}\right) - \Phi\left(\vec{y}^{\text{min}}\right)\right] \vec{x}_i + \Phi\left(\vec{y}^{\text{min}}\right)\text{ and }\\
    \vec{y}_i &= \vec{\mu} + \vec{\sigma}\sqrt{2}~\mathrm{erfinv}\left(2\vec{p}_i-1\right),
\end{align}
where $\Phi$ is the cumulative distribution function of the multivariate normal distribution with mean $\vec{\mu}$ and standard deviation $\vec{\sigma}$, and $\mathrm{erfinv}$ is the inverse error function. For each chemical parameter $j$, $\mu_j$ is taken from the nominal (standard) network, with nominal prefactors calculated using Eq.~\ref{eq:desprefacHH}, and $\sigma_j$ is determined according to Table~\ref{tab:uncertaintiesParameters}. 
\begin{table}
\caption{Sample widths for different studied parameters.}
\label{tab:uncertaintiesParameters}
\centering
\begin{tabular}{
    c 
    S[table-format = <4,table-text-alignment=left]@{~}
    l
    S[table-format = 3, table-text-alignment=right]@{~}
    l
    }
\hline\hline
Parameter type & \multicolumn{2}{c}{$\mu$} & \multicolumn{2}{c}{$\sigma$} \\
\hline
    Energy & {$< 200$} & K & {100} & K \\
           & {$\leq 1600$} & K & \multicolumn{2}{c}{$\frac{1}{2}\mu$} \\
           & {$> 1600$} & K & {800} & K \\
    Prefactor & \multicolumn{2}{c}{-} & {$10^2$} & s$^{-1}$ \\
\hline
\end{tabular}
\tablefoot{Energies are at least 0~K. Prefactors are varied in log-space.}
\end{table}
The bounds $y^{\text{min}}_j$ and $y^{\text{max}}_j$ are taken as $\min\left[0, \mu_j - \sigma_j\right]$ and $\mu_j+\sigma_j$, respectively. While approximate uncertainties are available for many gas-phase rate constants \citep{Millar2024}, this is not the case for the chemical parameters required for ice modeling. Thus, we chose reasonable values for their uncertainties, namely an uncertainty of 800~K for energies, which is a bit larger than ``chemical accuracy''\footnote{Chemical accuracy is the accuracy required to match or exceed experimental accuracy, and to make realistic chemical predictions. It is often taken to be 1~kcal~mol$^{-1}\approx500$~K.}. For a reaction with $\mu^{i+j}=1$~amu and a nominal reaction barrier of 800~K, the variation described in Table~\ref{tab:uncertaintiesParameters} leads to a variation of its reaction probability (Eq.~\ref{eq:reactionProbability}) of about 4 orders of magnitude at 10~K, and a nominal reaction barrier of 1600~K has a variation of its rate constant of about 5 orders of magnitude. Equation~\ref{eq:desprefacHH} underestimates the desorption prefactor by multiple orders of magnitude for many species \citep[see][]{Minissale2022}, so we chose a standard deviation of 2 orders of magnitude for all prefactors.

The amount of variation in the input parameters directly influences the magnitude of the variance in the abundances. Because we picked values for the input variation rather than take their uncertainties from, for example, literature, we are not able to do a proper error propagation and therefore cannot provide quantitative uncertainties. Nevertheless, the chosen variations in parameters are small enough that it is still possible to determine which species are qualitatively ``uncertain'', meaning highly sensitive to the input network. Because of the choice of $\sigma_j=0.5\mu_j$ if $\mu_j\leq1600$~K, species with a binding energy below 3200~K have their diffusion barriers varied between 25\% and 75\% of their nominal binding energies. Each of the 1000 $\vec{y}_i$ is considered a ``sample'' network and is used to replace the original network.

The code is set up such that the values of the energy barriers for reactions in surface and in bulk ice and corresponding chemical desorption reaction(s) are adjusted equally. Similarly, the binding energy and diffusion barrier of all bulk ice species change with those of water. Each of these 1000 samples is then run at the 225 different physical conditions. Running all models took less than 5 days on 110 cores of an AMD EPYC 7702P.

\subsection{Correlation coefficient}
\label{subsec:correlation_coefficient}
The correlation between the abundance of a species and a chemical parameter is calculated using the Pearson correlation coefficient, also referred to as Pearson's $r$. Both the abundances and parameters are first transformed using a rank-based inverse normal (RIN) transform \citep[rankit,][]{Bliss1967},
\begin{equation}
    f_{\mathrm{RIN}}\left(x_i\right) = \Phi^{-1}\left(\frac{\mathrm{R}\left[x_i\right]-0.5}{n}\right),
\end{equation}
where $n$ is the number of data points, and $\Phi^{-1}$ is the inverse normal cumulative distribution function. $\mathrm{R}\left[x_i\right]$ is the rank of $x_i$ in $\vec{x}$, such that the lowest value of $\vec{x}$ has $\mathrm{R}\left[x^{\mathrm{min}}\right]=1$ and its highest value has $\mathrm{R}\left[x^{\mathrm{max}}\right]=n$. Pearson's $r$ on the transformed data, which we will refer to as $r_{\mathrm{RIN}}$, is then calculated using SciPy 1.14.0 \citep{scipy}. $r_{\mathrm{RIN}}$ essentially functions as a Spearman correlation coefficient \citep{Spearman1904}, because it captures all monotonic correlations and not only linear correlations as would be the case without the rank transformation, is less sensitive to outliers and does not suffer in the case of non-normally distributed data. We use the RIN-transformed Pearson correlation coefficient instead of the Spearman correlation coefficient because it has more favorable statistical properties \citep{Bishara2012} and the implementation of the Pearson correlation coefficient in SciPy 1.14.0 is more complete. Previous works used the Pearson correlation coefficient without RIN transformation \citep[and instead used logarithmic abundances, see for example][]{Penteado2017,Iqbal2018,Furuya2022}, but from our testing that misses or underestimates various correlations.

The value of this coefficient indicates the ``strength'' of the correlation; a value of $1$ ($-1$) indicates a perfectly positive (negative) correlation, and values closer to 0 indicate a weaker correlation. The correlation coefficient itself does not give any information about the actual magnitude of the spread of abundances, only on the correlation between a certain parameter and the abundances relative to each other. Thus, interpreting whether a chemical parameter should be investigated more closely should always be done while keeping into account how big the spread of predicted abundances is. Without the RIN transformation, the Pearson correlation coefficient squared gives a measure of the explained variance. For example, a correlation coefficient of $0.5$ explains $0.5^2=0.25=25$\% of the variance in the samples. While this is not strictly true when using a RIN transformation, $r_{\mathrm{RIN}}^2\sigma^2$ still gives a general idea of both the variance of the abundances and the strength of the correlation.

There are two sources of error in the determination of the correlation coefficient. Firstly, a sampling error as a result of the finite number of Monte Carlo samples. From a convergence test with 2000 samples at the extremes of densities and temperatures in our grid, the 95\% confidence interval for this error is around $\pm 0.08$. Generating the samples using Latin Hypercube Sampling or Sobol' sequences \citep{Sobol1967} did not result in better convergence. The second error, from the correlation coefficient itself, is calculated using bootstrapping with the BC\textsubscript{a} method \citep[][chap.~14]{Efron1994} with 2000 samples. The two errors are assumed to be independent, such that the total confidence interval is their sum. This is very likely an upper limit of the width of the confidence interval (maybe even to a factor of 2), since the correlation coefficients calculated with the convergence test also include the error from the correlation coefficient itself. Only correlations with $p\leq0.05$ are reported and discussed.

In the spirit of open science, we made all code to generate the data and create the figures shown here publicly available under the GNU GPLv3 license\footnote{\url{https://github.com/uclchem/SensitivityAnalysisIces}}. All data at $\zeta=1.3\times10^{-17}$~s$^{-1}$ and $F_{\mathrm{UV}}=1$~Habing is available on Zenodo\footnote{\url{https://doi.org/10.5281/zenodo.17463693}}.

\section{Results and discussion}
\label{sec:Results}
\subsection{Understanding the correlation coefficient}
\label{subsec:understanding_the_correlation_coeff}
\begin{figure}
  \capstart
  \resizebox{\hsize}{!}{\includegraphics{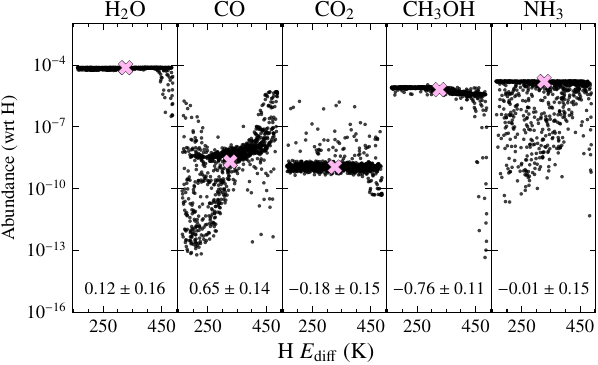}}
  \caption{Abundances of all samples at $T=10$~K, $n_{\text{H}}=10^5$~cm$^{-3}$, $\zeta=1.3\times10^{-17}$~s$^{-1}$ and $F_{\mathrm{UV}} = 1$~Habing at $10^4$~years as function of the hydrogen diffusion barrier. Numbers at the bottom indicate $r_{\mathrm{RIN}}$ and its 95\% confidence intervals. Pink crosses indicate nominal abundances and hydrogen diffusion barrier.}
  \label{fig:exampleCorrelationsWithHdiff}
\end{figure}

Figure~\ref{fig:exampleCorrelationsWithHdiff} shows the abundances of a few major ice species at $T=10$~K, $n_{\text{H}}=10^5$~cm$^{-3}$, $\zeta=1.3\times10^{-17}$~s$^{-1}$ and $F_{\mathrm{UV}} = 1$~Habing at $10^4$~years as a function of the sampled hydrogen diffusion barrier. Here, it can be seen that the \ch{H2O} abundance as a function of the hydrogen diffusion barrier is a flat line, meaning that they are not correlated. On the other hand, there is a strong correlation between the \ch{CO} abundance and \ch{H} diffusion barrier, and this is indeed reflected in the value of $r_{\mathrm{RIN}}$.

Methanol, \ch{CH3OH}, also shows a strong correlation, where the abundance at high H diffusion barriers ($>425$~K) decreases by nearly 10 orders of magnitude. This is an example of a nonlinear correlation, also when using logarithmic abundances. As mentioned in Sect.~\ref{subsec:correlation_coefficient}, $r$ without the RIN transformation underestimates the strength of this correlation, with $r=-0.4\pm0.11$. This highlights the need to use a rank-based correlation coefficient.

\ch{H2O}, \ch{CH3OH} and \ch{NH3} abundances all show a sharp decrease at H diffusion barriers above 425~K. At these high diffusion barriers, H becomes stationary, which leads to the unrealistic build-up of H ice up to an abundance of $10^{-4}$, which corresponds to about $10^2$~monolayers. All carbon-, oxygen-, or nitrogen-bearing molecules are pushed to the bulk, where reactions are extremely slow due to the necessary diffusion of reagents through \ch{H2O}. Thus, the abundance of larger species that rely on their formation on ices sharply decreases.

\begin{figure}
  \capstart
  \resizebox{\hsize}{!}{\includegraphics{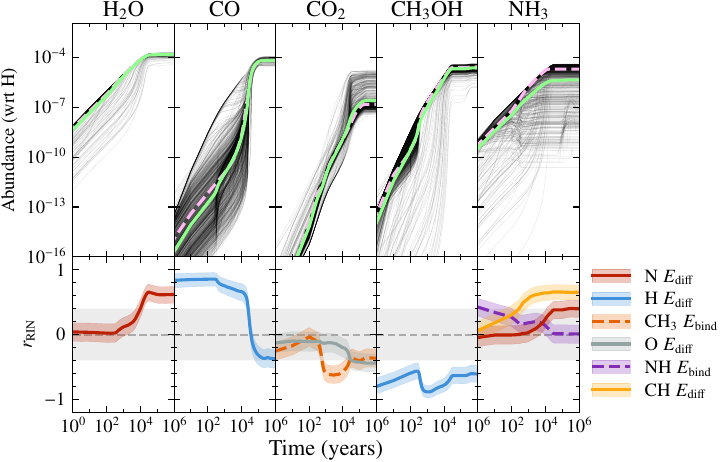}}
  \caption{Abundances (top) and correlation coefficients (bottom) of \ch{H2O}, \ch{CO}, \ch{CO2}, \ch{CH3OH}, and \ch{NH3} ices at $T=10$~K, $n_{\text{H}}=10^5$~cm$^{-3}$, $\zeta=1.3\times10^{-17}$~s$^{-1}$ and $F_{\mathrm{UV}} = 1$~Habing. In the top panels, the dashed pink line indicates the abundance using the nominal network, black lines indicate the individual samples, and light green is the log-average of all samples. In bottom subplots, the dashed gray line indicates perfectly uncorrelated parameters ($r_{\mathrm{RIN}}=0$), and the filled gray area indicates ``weakly'' correlated parameters ($\left|r_{\mathrm{RIN}}\right|<0.4$). The colored filled areas correspond to the 95\% confidence intervals of the correlation coefficients.}
  \label{fig:10K_1e5cm-3_1.0_1.0_sensitivities}
\end{figure}

To illustrate the analysis workflow, we first show results for one set of physical conditions. The top panels in Fig.~\ref{fig:10K_1e5cm-3_1.0_1.0_sensitivities} present the abundances of five major ice species as function of time at $T=10$~K, $n_{\text{H}}=10^5$~cm$^{-3}$, $\zeta=1.3\times10^{-17}$~s$^{-1}$ and $F_{\mathrm{UV}} = 1$~Habing for all 1000 samples in black. The dashed pink lines indicate the abundance progression using the nominal network, whereas the light green lines show the log-average (or geometric mean) of all samples. All ``strong'' correlations, meaning $\left|r_{\mathrm{RIN}}\right|\geq0.4$, at any point in time are plotted in the bottom panels.

There is a considerable difference in the sensitivities before and after \ch{CO} freeze-out, which occurs at around $10^4$~years at this density. For example, the \ch{H2O} abundance is first not sensitive to any parameter, however at $10^4$~years becomes very sensitive to the \ch{N} diffusion barrier.

It might not be immediately obvious why the \ch{H2O} ice abundance is sensitive to the \ch{N} diffusion barrier. Water and ammonia are mostly formed by multiple barrierless hydrogenations of \ch{O} and \ch{N} atoms, respectively. The nominal binding energies of atomic \ch{H} and \ch{N} are 650 and 720~K, respectively, meaning that the corresponding nominal diffusion barriers are 325 and 360~K. The main destruction pathways of atomic oxygen are through barrierless \ch{N + O -> NO} and \ch{O + H -> OH}. At later times, the gaseous O abundance depletes, because most of the oxygen is locked up in water already on the grain. Thus, the freeze-out rate of oxygen decreases, and the abundance of surface abundance of \ch{O} decreases from $\sim$$10^{-6}$ to $\sim$$6\times10^{-9}$ with respect to hydrogen nuclei. This also happens because CO freeze-out pushes surface O to the bulk, where diffusion is much slower. Therefore, the very limited amount of surface \ch{O} is used by the two reactions mentioned above, such that a more efficient \ch{N} diffusion further depletes the amount of surface oxygen, and decreases the formation of \ch{OH} and thus \ch{H2O}. Additionally, with most of the \ch{CO} frozen out, atomic H is consumed much faster by \ch{H + CO -> HCO}. This means that the two reactions mentioned above will swap in importance, that is to say, the ratio of the two reaction rates, $\left(k_{\text{reac}}^{\text{N} + \text{O}} X_{\text{N}}\right)/\left(k_{\text{reac}}^{\text{H} + \text{O}}X_{\text{H}}\right)$, goes up, and so the diffusion barrier of \ch{N} becomes more important as well.

The opposite effect can be seen for the abundance of \ch{CO} ice, where it is very sensitive to the hydrogen diffusion barrier at early times, and at later times that sensitivity is much weaker. CO on the grain is mainly formed and destroyed by \ch{HCO + H -> CO + H2} and \ch{CO + H -> HCO}, respectively. A reaction between \ch{HCO} and \ch{H} can also form \ch{H2CO}. At early times, when \ch{HCO} forms and subsequently reacts with \ch{H}, the abundance of \ch{CO} will go down overall, because some will form \ch{H2CO}. A higher diffusion barrier of atomic hydrogen will lead to a lower formation rate of \ch{HCO}, and thus a higher abundance of \ch{CO} ($r_{\mathrm{RIN}}\approx0.9$). At later times, however, the \ch{H} abundance is so low that the correlation disappears. Methanol is still sensitive to the \ch{H} diffusion barrier, because the abundance of methanol remains unchanged as all species are pushed to the bulk at $10^4$~years and then reaches steady state due to the slow reactions in bulk ice.

\ch{NH3} is sensitive to the \ch{NH} binding energy at early times. In the nominal network, the \ch{NH} binding energy is 2600~K, which means that thermal desorption is negligible at 10~K. However, the barrierless reaction \ch{N + H -> NH} is quite exothermic ($\Delta H_{\mathrm{reac}}=-88.43$~kcal~mol$^{-1}$), and at these early times there is not yet a full monolayer of ice on the dust grains. Thus, $\eta_{\mathrm{CD,grain}}^{\text{NH}}\approx 0.56$. Decreasing and increasing the binding energy of NH by 800~K changes the desorption probabilities to about 0.67 and 0.47, respectively, so this will have a large effect on the \ch{NH} abundance, and in turn on the abundance of \ch{NH3}. This sensitivity disappears at later times, because the total ice abundance increases and the chemical desorption probability goes down. At later times, \ch{NH3} is sensitive to the \ch{CH} diffusion barrier at many physical conditions because of the barrierless reaction \ch{NH3 + CH -> CH2NH2}, the main destruction pathway of \ch{NH3}. The rate constant of this reaction, which is determined by the diffusion barrier of \ch{CH}, imposes an upper limit on the \ch{NH3} abundance.

\subsection{Across physical conditions}
\label{subsec:across_physical_conditions}
For the species of interest here, the correlations are relatively unaltered by changing the cosmic ray ionization rate \citep[though this might not be the case at even higher rates, such as in the Central Molecular Zone;][]{LePetit2016}. That is not to say that the abundances are unchanged by increasing the cosmic ray ionization rate, but the sensitivities are. Thus, for simplicity, we will continue discussing the abundance profiles and sensitivities using the standard rate of $\zeta=1.3\times10^{-17}$~s$^{-1}$. Similarly, the UV field strength has no effect on the ice abundances for densities above $10^{4}$~cm$^{-3}$ and no effect on the sensitivities at all physical conditions, so we will use 1~Habing for the rest of this work. 

\begin{figure*}
     \centering
     \capstart
     \includegraphics[width=18cm]{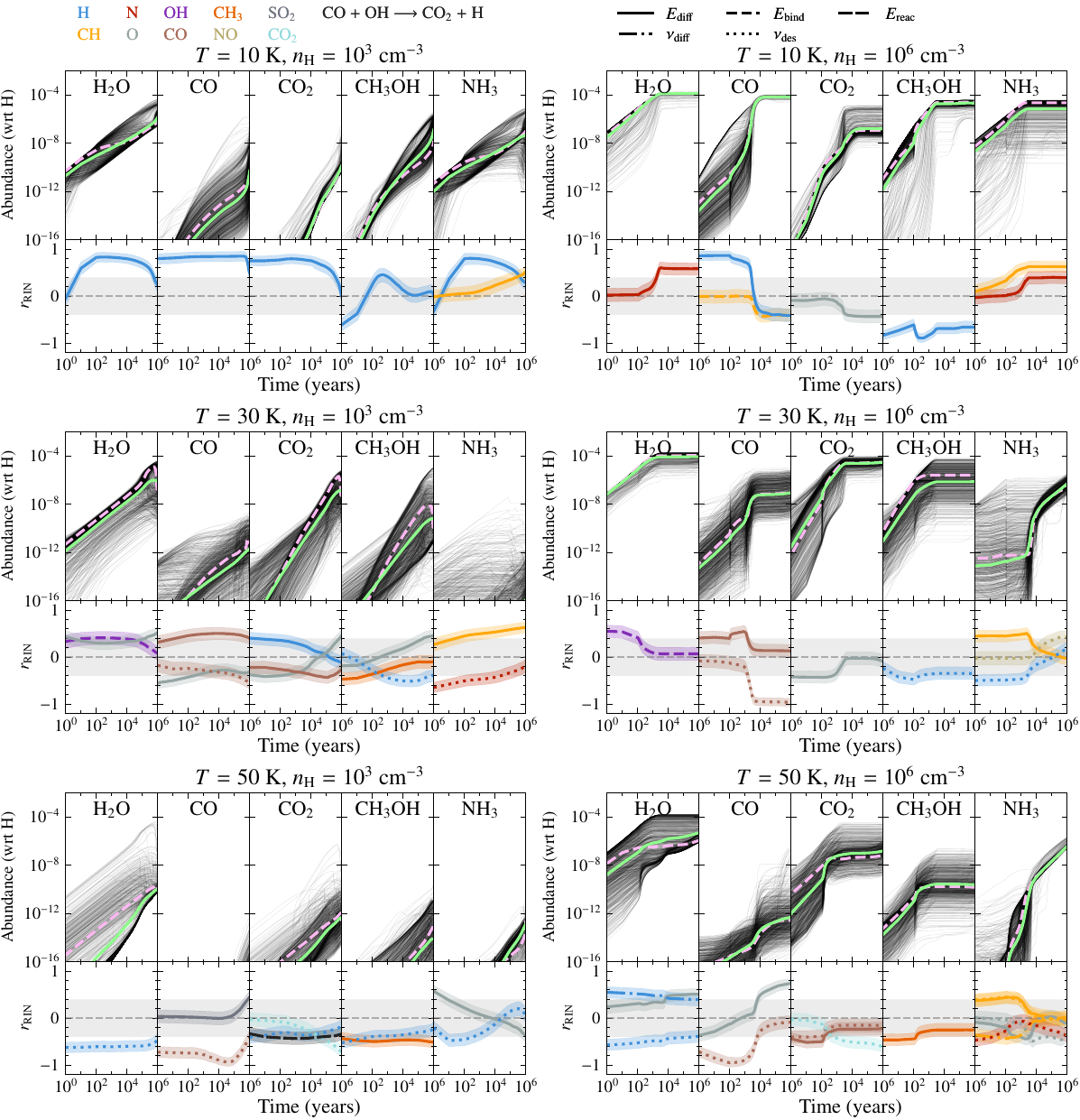}
     \caption{Abundances (top) and correlation coefficients (bottom) of \ch{H2O}, \ch{CO}, \ch{CO2}, \ch{CH3OH}, and \ch{NH3} ices at various physical conditions indicated by the text above each subfigure. In the top panels, the dashed pink line indicates the abundance using the nominal network, black lines indicate the individual samples, and light green is the log-average of all samples. In bottom subplots, the dashed gray line indicates perfectly uncorrelated parameters ($r_{\mathrm{RIN}}=0$), and the filled gray area indicates ``weakly'' correlated parameters ($\left|r_{\mathrm{RIN}}\right|<0.4$). The color indicates which species or reaction the parameter belongs to, and the linestyle indicates its type. The colored filled areas correspond to the 95\% confidence intervals of the correlation coefficients. Legends are the same for all panels.}
     \label{fig:VariousPhysicalConds}
\end{figure*}

Figure~\ref{fig:VariousPhysicalConds} shows sensitivities for various physical conditions, as indicated above each set of panels. The temperature changes across the rows, whereas the density changes between the two columns. As expected, we observe that an increase in density mainly decreases the chemical timescale, and that the correlations are only affected to a limited extent. This can be seen by, for example, comparing Fig.~\ref{fig:10K_1e5cm-3_1.0_1.0_sensitivities} at $n_{\mathrm{H}}=10^{5}$~cm$^{-3}$ and the top-right panel of Fig.~\ref{fig:VariousPhysicalConds} at $n_{\mathrm{H}}=10^{6}$~cm$^{-3}$. On the other hand, changing the temperature allows other reactions to take over, such that each temperature has different sensitivities. For clarity, only 10, 30 and 50~K are shown in Fig.~\ref{fig:VariousPhysicalConds}.

At 10~K and $10^3$~cm$^{-3}$ (top left in Fig.~\ref{fig:VariousPhysicalConds}), many species are sensitive to the hydrogen diffusion barrier. There is no catastrophic CO freeze-out at this low density. This means that the H abundance on the surface is high, so species are still sensitive to its diffusion barrier. On the other hand, at higher densities, \ch{CO} freeze-out depletes the H abundance, as mentioned in Sect.~\ref{subsec:understanding_the_correlation_coeff}. The diffusion barriers of most species are too high for them to diffuse, so most species are stationary at this temperature. Small variations in the H diffusion barrier lead to large differences in the rate constants. For example, variation of the diffusion barrier between 350 and 450 K gives a difference in the rate constant of a factor $\sim$$2 \times 10^4$ (for the same prefactor) at 10~K. At 30~K, the same variation only results in a difference of a factor of 28. On the other hand, the variation of prefactors results in a factor of $10^4$, which is why the abundances are more sensitive to the prefactors at 30~K.

The abundance of many species at 30~K and low density are correlated with the diffusion barrier of atomic oxygen. Since \ch{O} is the more mobile reactant between \ch{C} and \ch{O}, some carbon- and oxygen-bearing species, like \ch{CO2}, rely on the diffusion of \ch{O} to be formed on the ice. For example, the second most important formation reaction of \ch{CO2} at 30~K and $n_{\mathrm{H}}=10^3$~cm$^{-3}$ is \ch{H2CO + O -> CO2 + H2}. Thus, a smaller diffusion barrier of \ch{O} leads to a larger abundance of these species ($r_{\mathrm{RIN}}<-0.4$ at early times). On the other hand, the formation of water occurs mainly through diffusion of \ch{H} atoms, and lowering the diffusion of \ch{O} atoms depletes the amount of oxygen (by formation of the species mentioned above) and thus lowers the \ch{H2O} abundance ($r_{\mathrm{RIN}}>0.4$).

The \ch{CO} abundance at 30 and 50~K and \ch{CO2} abundance at 50~K indicate the need for the use of accurate values for the desorption prefactor $\nu_{\mathrm{des}}$. These are highly sensitive to the desorption prefactor, as indicated in the middle and bottom panels of Fig.~\ref{fig:VariousPhysicalConds}, with a spread of about 4 orders of magnitude. This shows that the choice of prefactor is crucial, and astrochemical models need to use new chemical insight to calculate them \citep{Minissale2022,Ligterink2023}.

An interesting observation from Fig.~\ref{fig:VariousPhysicalConds} is that only one reaction energy barrier appears in these graphs, \ch{CO + OH -> CO2 + H} ($E_{\text{reac}}=1000$~K). The maximum strength of correlation with this reaction is still quite weak, with $r_{\text{RIN}}\approx-0.45$. This means that all other reaction barriers have correlations $\left|r_{\mathrm{RIN}}\right| < 0.4$ with all five species at all times and all of these six physical conditions. To confirm the insensitivity of abundances to reaction energy barriers, we repeated the calculations with only varying reaction energy barriers and keeping all other chemical parameters (binding energies, diffusion barriers, and prefactors) fixed. The results of these calculations are shown in App.~\ref{app:variation_of_reaction_energy_barriers}. The analysis here is limited to the total ice abundances, and the picture might be different if this is further differentiated to the surface and bulk abundances. The lack of correlations with reaction energy barriers is further discussed in Sect.~\ref{subsec:theory_and_exp}.

\subsection{Parameters showing strongest correlation}
Many of the same parameters appear in Fig.~\ref{fig:VariousPhysicalConds}. We have summarized these in Table~\ref{tab:parametersToInvestigate}. The references given in this table are by no means an exhaustive list (especially for \ch{H} diffusion), but are there to give the reader an entry point into further reading.
\begin{table*}
\caption{Selection of chemical parameters that many species are correlated with, and the temperature(s) at which they were found to be have strong correlations with ice abundances.}
\label{tab:parametersToInvestigate}
\centering
\begin{tabular}{l l l@{ }l l l l}
\hline\hline
Species & Type & \multicolumn{2}{c}{Nominal value} & Temperature (K) & Relevant literature \\
\hline
    \ch{H} & $E_{\mathrm{diff}}$ & 325 & K & 10 & \citet{Asgeirsson2017, Senevirathne2017,Hama2012} \\
     & $\nu_{\mathrm{des}}$ & $4.4\times10^{12}$ & s$^{-1}$ & 30\textendash50 & \citet{Minissale2022} \\
    \ch{N} & $E_{\mathrm{diff}}$ & 360 & K & 10 & \citet{Zaverkin2021} \\
    \ch{O} & $E_{\mathrm{diff}}$ & 800 & K & 10\textendash50 & \citet{Pezzella2018} \\
    \ch{CO} & $E_{\mathrm{diff}}$ & 650 & K & 30 & \citet{Karssemeijer2013,Kouchi2020} \\
      & $\nu_{\mathrm{des}}$ & $1.1\times10^{12}$ & s$^{-1}$ & 30\textendash50 & \citet{Minissale2022} \\
    \ch{HCO} & $E_{\mathrm{diff}}$ & 1200 & K & 10\textendash50 & \citet{Ferrero2020}\tablefootmark{a}, \citet{Bovolenta2022}\tablefootmark{a} \\
    \ch{CH3} & $E_{\mathrm{diff}}$ & 800 & K & 10\textendash50 & \citet{EnriqueRomero2025} \\
\hline
\end{tabular}
\tablefoot{Nominal diffusion barriers are half of the binding energy taken from \citet{Wakelam2017}. Nominal prefactors are calculated according to Eq.~\ref{eq:desprefacHH} \citep{Hasegawa1992}.
    \tablefoottext{a}{Determines the binding energy.}
}
\end{table*}

We can compare our main correlations at 10~K with those of previous work. Comparing our results with those of \citet{Penteado2017}, we find quite similar results. They find also the binding energy of \ch{C} to be relevant, though they use a nominal binding energy of 715~K, and here a binding energy of 10000~K corresponding to chemisorbed \ch{C-H2O} is used. They find \ch{CH2} diffusion to be sensitive as well, but our results show \ch{CH} is actually the more sensitive one due to the slightly higher values we use here for their binding energies. We do not have any correlation with the \ch{H2} binding energy, because we incorporate an encounter desorption (Sect. \ref{subsec:encounter_desorption}) method already. \citet{Iqbal2018} and \citet{Furuya2022} also found the diffusion barriers of H, N, HCO, and O to constrain the predicted abundances. Instead of \ch{CH}, they found the \ch{CH2} diffusion barrier, however that is also because the network they use makes \ch{CH2} more mobile than \ch{CH}, whereas our network has the opposite behavior. \citet{Sil2024} calculated the binding energy of \ch{CH} on \ch{H2O} to be 2400~K, but \citet{Wakelam2017} found that \ch{CH} barrierlessly reacts with \ch{H2O} at a lower level of theory. The binding energy of \ch{CH2} was estimated to be 1400~K by \citet{Wakelam2017}.

Of the parameters listed in Table~\ref{tab:parametersToInvestigate}, the H diffusion barrier has been investigated the most. \citet{Asgeirsson2017} and \citet{Senevirathne2017} both investigated the diffusion of atomic hydrogen on crystalline and amorphous solid water using the same potential energy surface. They found diffusion to binding energy ratios of 0.64 and 0.37 respectively. This large difference is likely due to different surface structures. \citet{AlHalabi2007} also investigated diffusion of \ch{H}, and directly obtain the diffusion coefficient from molecular dynamics simulations.

The diffusion of atomic nitrogen on amorphous solid water was studied recently by \citet{Zaverkin2021}, who found a ratio of 0.76. \citet{Pezzella2018} showed that the average diffusion barrier of atomic oxygen is 275~K, corresponding to a diffusion to binding energy ratio of around 0.3 using their binding energy. This confirms that there is no universal value for this ratio \citet{Furuya2022}, not even for atomic species, as is used in most astrochemical models.

The diffusion barrier of \ch{HCO} has not been studied explicitly. However, the binding energies of HCO reported in literature are larger than the ones used in this work. Thus, the diffusion of HCO is likely not as relevant as it appears from this study. We use a binding energy of 2400~K, as reported by \citet{Ferrero2020}, whereas \citet{Bovolenta2022} reports a binding energy of around 1400~K. This large difference can likely be attributed to the different clusters, as the former uses a cluster with a cavity, and the latter uses a spherical cluster of only 12\textendash15 water molecules. Because \ch{HCO} is formed from \ch{CO}, it is most likely formed on a \ch{CO}-rich surface. The strength of interaction with a \ch{CO} surface is lower than on \ch{H2O}, meaning \ch{HCO} diffuses more rapidly. To the best of our knowledge, the binding or diffusion of \ch{HCO} on \ch{CO} has not yet been studied.

The diffusion of \ch{CH3} was studied recently by \citet{EnriqueRomero2025} on a small cluster of 14 \ch{H2O} molecules. Due to the small cluster size, the binding energies and diffusion barriers can be best regarded as lower limits. Because they focused on the formation of \ch{CH3CN} arising from short-range diffusion, the long-range diffusion of \ch{CH3} was not investigated. 

\subsection{Limitations}
\label{subsec:limitations}
Nondiffusive chemistry means that reactions can take place between any two reactants in proximity. For example, after a reaction has taken place, the product(s) of that reaction can again react with other close-by species. This means that diffusion of heavy radicals is not necessary to produce larger molecules such as complex organic molecules, and thus allows for their formation at lower temperatures. Diffusion is still necessary for the formation of the radicals. This is inherently present in kinetic Monte Carlo simulations, which can reproduce laboratory experiments at low temperatures where diffusion of larger radicals is slow \citep{Fedoseev2015, Fedoseev2017, Simons2020,Ioppolo2020}. It has recently been implemented into rate equation models \citep{Jin2020}, but was not considered in this work. We expect that if it were, the correlations of abundances with binding energies of weakly bound species would increase because the nondiffusive chemistry would lead to cyclic H abstraction and addition, making chemical desorption more efficient \citep[see Sect~4.1 of][]{Jin2020}. This also means that the abundances would become more sensitive to how chemical desorption is treated. On the other hand, we expect that correlations with diffusion barriers of larger radicals (e.g.,~\ch{HCO} and \ch{CH3}) to become less important because as a radical is formed, it can react with other radicals without needing to overcome diffusion barriers. As shown in Fig.~9 of \citet{Borshcheva2025}, the molecular diffusion barriers indeed matter less in the nondiffusive framework. However, larger molecules such as complex organic molecules are likely most affected by this. Occupation of deeper binding sites by other species can also greatly speed up diffusion \citep[see for example][]{Karssemeijer2013,Zaverkin2021} and is not taken into account in current astrochemical models.

Besides Langmuir-Hinshelwood reactions, chemistry on ices can also occur through Eley-Rideal reactions. Eley-Rideal reactions can occur when a species from the gas-phase adsorbs on the surface of the ice next to or on top of an already adsorbed species. In the network used here, there is a very limited number of Eley-Rideal reactions, meaning that chemistry must be driven by Langmuir-Hinshelwood (diffusive) reactions. Including more Eley-Rideal reactions might increase the importance of energy barriers. However, because Eley-Rideal reactions rely on collisions with the adsorbates, this is likely most important for major ice species like \ch{CO}.

Varying the reaction energy barrier affects the tunneling probability, and thus the reaction rate. However, we note that the tunneling probability estimated with the rectangular barrier approximation can be off by orders of magnitude compared to quantum chemical instanton calculations. For example, using Eq.~1 from \citet{Lamberts2017} and data there-in, we find an instanton tunneling probability of $\sim$$4\times10^{-9}$ at 10~K for \ch{CH4 + OH -> CH3 + H2O}. Using the energy barrier they calculated using density functional theory (2575~K) and the rectangular barrier approximation, we find $P_{\text{reac}}\approx3\times10^{-13}$. Decreasing this barrier by 800~K increases the probability to $\sim$$4\times10^{-11}$, still two orders of magnitude off from the instanton probability. An ``effective`` energy barrier, the energy barrier that results in a correct tunneling probability from Eq.~\ref{eq:reactionProbability}, or Model~3 suggested in \citet{Zheng2010} for correct temperature dependence could be used, but for many reactions no tunneling probabilities are available.

\section{Implications}
\label{sec:implications}
\subsection{Poorly constrained species}
\label{subsec:observations}
\begin{figure}
  \capstart
  \resizebox{\hsize}{!}{\includegraphics{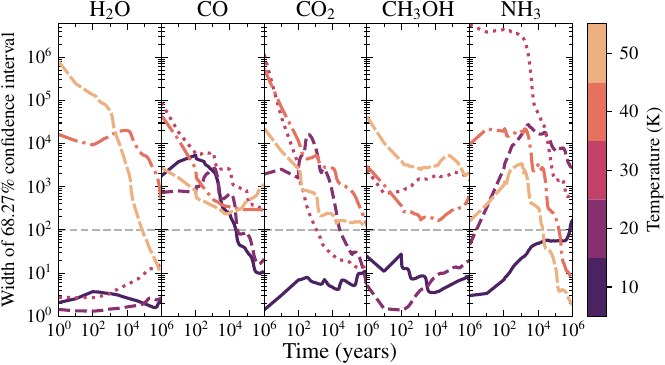}}
  \caption{Width of the 68.27\% confidence intervals of abundances of selected ice species at various temperatures (averaged over all densities) over time at $\zeta=1.3\times10^{-17}$~s$^{-1}$ and $F_{\mathrm{UV}}=1$~Habing. The solid, dashed, dotted, dashed-dotted, and densely dotted lines correspond to 10, 20, 30, 40 and 50~K, respectively. The dashed gray line indicates a width of 2 orders of magnitude.}
  \label{fig:confidence_intervals_abundances}
\end{figure}

As mentioned in Sect.~\ref{subsec:sampling_chemical_parameters}, we are not able to give quantitative uncertainties of the abundances. Nevertheless, qualitative uncertainties are still very beneficial to properly constrain physical parameters of observed clouds. At 10~K, the abundance of \ch{H2O} is quite insensitive to changes in the ice chemistry, such that observations of \ch{H2O} ice can be accurately fitted. On the other hand, the abundances of \ch{CO} and \ch{CH3OH} are very sensitive to changes in the network (see for example the top row of Fig.~\ref{fig:VariousPhysicalConds}), such that fits of observations are also sensitive to changes in the network. We note again that the hydrogen diffusion barrier is varied with 162.5~K from the nominal value in either direction, and this is a similar change as going from $E_{\mathrm{diff}}^{\mathrm{H}}=0.25E_{\mathrm{bind}}^{\mathrm{H}}$ to $E_{\mathrm{diff}}^{\mathrm{H}}=0.75E_{\mathrm{bind}}^{\mathrm{H}}$. This small (in terms of energy) change in hydrogen diffusion barrier has an extremely large effect on many species at 10~K. At 20~K and above, variations of diffusion barriers of larger species start to strongly affect the predicted abundances. Figure~\ref{fig:confidence_intervals_abundances} shows the width, meaning the ratio of the upper limit to the lower limit, of the 68.27\% confidence interval of abundances at various temperatures, averaged over all densities. In general, the width goes down at later times, because these species are mostly consumed by reactions. It is clear that the abundances are often uncertain to within 2\textendash4 orders of magnitude.

This shows the risk of extracting chemical parameters from comparisons between models and observations, such as a diffusion barrier to binding energy ratio \citep{Garrod2011,Borshcheva2025}. From our sensitivity analysis, we conclude that observations of specific species cannot determine a ratio $\chi$ that is universal for all atoms and molecules.

\subsection{Assumptions in models}
\label{subsec:models}
Astrochemically relevant ices (e.g.~amorphous solid water) do not have unique binding sites, but have a distribution of binding sites with various depths and probabilities \citep{Amiaud2006,He2016,Minissale2022}. Early attempts to include this in astrochemical models relied on treating the species in different binding sites as entirely different species in the network (i.e.~with separate abundances and reactions), with rates connecting them \citep{Cuppen2011,Grassi2020}. This made it too computationally expensive to do so for every species. \citet{Furuya2024} developed an efficient method to include binding energy distributions. While we do not take binding energy distributions into account, we can still probe whether a species will be strongly affected by including distributions. It is important that the binding energy distributions are well-studied and implemented, as simply assuming a normal distribution is not valid for many species \citep[see for example][]{Bovolenta2022,EnriqueRomero2024}, and even the assumed width of the distribution has an effect on the abundances \citep[see Figs.~6 and 7 of][]{Furuya2024}. Species that are strongly correlated with binding energies or diffusion barriers will be affected by the assumed distributions.

Recently, there have been multiple works that vary the binding energy, and as a result the diffusion barrier, of species by the amount of CO in the ice \citep[see for example][]{Molpeceres2024}. This becomes especially important after the catastrophic \ch{CO} freeze-out has occurred, which greatly increases the abundance of \ch{CO} on the surface, thus decreasing the interaction strength of adsorbates with the ice surface. Some works simply assume ratios of the interaction strengths $E_{\text{bind on CO}}^i/E_{\text{bind on \ch{H2O}}}^i$ for all species \citep[see for example][]{Kalvans2024}, though this should be done with care because, as shown above, assumptions about diffusion barriers greatly influence the calculated abundances. Thus, if scaling the binding energy by the amount of \ch{CO}, we recommend the method by \citet{Molpeceres2024}, where you calculate the binding energies of species of interest on \ch{CO} using density functional theory and verify that varying the binding energy on \ch{CO} in the model of other species does not affect abundances.

\subsection{Recommendations for experiments and calculations}
\label{subsec:theory_and_exp}
One of the main findings is that the calculated ice abundances of many species are not sensitive to reaction energy barriers. That means that getting accurate reaction energy barriers (using e.g.,~high-level density functional theory) is often not necessary. This also applies to complex organic molecules, as discussed in App.~\ref{app:complexOrganicMolecules}.

This is because of the treatment of competition. Consider a reaction between $i$ and $j$, of which $i$ is the one with the lower diffusion barrier. At low diffusion barriers, the diffusion rate ($k_{\mathrm{diff}}^{i}$, Eq.~\ref{eq:hoppingRate}) becomes much larger than the reaction rate $\kappa^{i + j}$ if the reaction has a barrier. Thus, the fraction describing competition (Eq.~\ref{eq:competitionFraction}) can be approximated as $\kappa^{i + j}/k_{\mathrm{diff}}^{i}$, and the resulting reaction rate constant (Eq.~\ref{eq:rateConstantInclCompetition}) becomes approximately
\begin{equation}
    k_{\mathrm{reac}}^{i + j} \propto \frac{\kappa^{i + j}}{k_{\mathrm{diff}}^{i}}k_{\mathrm{diff}}^{i}=\kappa^{i + j},
    \label{eq:fastDiffusionLimit}
\end{equation}
independent of the diffusion rate. On the other hand, reactions with a low energy barrier or for which the diffusion barriers of the reagents are high have $\kappa^{i + j}\gg k_{\mathrm{diff}}^{i}$, such that
\begin{equation}
    k_{\mathrm{reac}}^{i + j} \propto \frac{\kappa^{i + j}}{\kappa^{i + j}}k_{\mathrm{diff}}^{i}=k_{\mathrm{diff}}^{i},
    \label{eq:slowDiffusionLimit}
\end{equation}
independent of $\kappa^{i + j}$ and thus the reaction barrier height. In other words, the reaction rate constant $k_{\text{reac}}^{i+j}$ is proportional to the rate of the slowest process.

\begin{figure}
  \capstart
  \resizebox{\hsize}{!}{\includegraphics{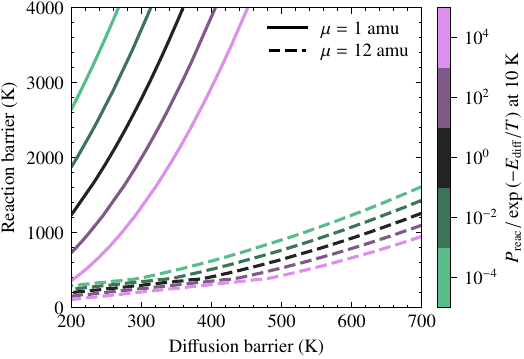}}
  \caption{Ratio between the reaction probability (Eq.~\ref{eq:reactionProbability}) and the Boltzmann factor for diffusion at 10~K, for reactions with different energy barriers and diffusion barrier of the most mobile reagent. Solid lines indicate a reaction where tunneling is efficient ($\mu=1$~amu), and dashed lines indicate a reaction where tunneling is inefficient ($\mu=12$~amu).}
  \label{fig:ratioReactionDiffusion_10K}
\end{figure}

Figure~\ref{fig:ratioReactionDiffusion_10K} shows the ratio between the reaction probability (Eq.~\ref{eq:reactionProbability}) and the Boltzmann factor for diffusion at 10~K. Low ratios, indicated by green, correspond to the fast diffusion limit (Eq.~\ref{eq:fastDiffusionLimit}) and high ratios, indicated by purple, correspond to the slow diffusion limit (Eq.~\ref{eq:slowDiffusionLimit}). We neglect the influence of the prefactors in this figure, meaning that we show the ratio of the reaction probability to the diffusion probability, and not their respective rate constants. The exact values of the tunneling mass $\mu$ in Fig.~\ref{fig:ratioReactionDiffusion_10K} are not important, but two examples are shown to indicate how the tunneling efficiency affects the trend. For reactions where tunneling through the reaction energy barrier is inefficient, like if a carbon atom is being exchanged ($\mu=12$~amu, shown in Fig.~\ref{fig:ratioReactionDiffusion_10K}), the height of the reaction energy barrier affects the ratio more. 
Many relevant reactions on ices, especially at this low temperature, have a reaction barrier below 1000~K to 2000~K, tunneling mass of $\mu=1$~amu and reagents with diffusion barriers above 300~K. Thus, they fall in the slow-diffusion limit, where $P_{\mathrm{reac}}\gg \exp \left(-E_{\mathrm{diff}}/T\right)$, and their reaction rates are dictated mainly by the rate of diffusion of the most mobile species according to Eq.~\ref{eq:slowDiffusionLimit}. This means that an accurate reaction probability, calculated with for example instanton theory \citep[e.g.~in][]{Lamberts2017}, or an exact energy barrier for the reaction is not necessary. On the other hand, if there is reason to suspect that the reaction might occur in the fast-diffusion limit (green in Fig.~\ref{fig:ratioReactionDiffusion_10K}), for example because it has a very high barrier or involves movement of relatively heavy atoms making tunneling inefficient, then the actual reaction probability becomes more important. However, these reactions are generally less important for the ice chemistry. At higher temperatures these regimes shift further down and right, because diffusion is a temperature-dependent process whereas the reactions are often tunneling-dominated and so temperature-independent. 

To give an example, the \ch{CO2} abundance at 50~K correlates with the reaction barrier of \ch{CO + OH -> CO2 + H}, as can be seen in the bottom-left panel of Fig.~\ref{fig:VariousPhysicalConds}. At the nominal CO and OH diffusion barriers and energy barrier (1000~K), the ratio between the sum of the diffusion rate constants and the reaction rate constant is around $2\times10^{-3}$, in the fast diffusion regime. On the other hand, at the minimum of the range of its energy barrier, the ratio becomes around 50, in the slow diffusion regime. Thus, the height of the reaction barrier is important for the abundance of \ch{CO2}.

This means that once the diffusion barriers are properly constrained, accurate reaction energy barriers are often not required to have robust chemical models. After an approximate calculation (using e.g.~a lower accuracy density functional or machine-learned interatomic potential) or measurement has been done, it can be decided if it needs closer investigation. If the reaction falls in the slow diffusion regime at the physical conditions the reactants are expected to be available, it is probably not needed to obtain a more accurate energy barrier. On the other hand, if the reaction falls close to the switch or in the fast diffusion regime, a more accurate energy barrier might be necessary. We note again that this is a diffusive rate equation model that has no microscopic detail. In a model with more microscopic detail, for example kinetic Monte Carlo, where explicit competition between different reactions is taken into account, the situation might be different.

\section{Summary and conclusions}
\label{sec:summary_and_conclusions}
We investigated the correlation of abundances of ice species with binding energies, diffusion barriers, reaction energy barriers, and desorption and diffusion prefactors. We also improved the treatment of interstellar ice chemistry in UCLCHEM by updating the chemical desorption formalism, limiting the amount of \ch{H2} on the surface, including better calculation of the tunneling probability, and correctly calculating the desorption prefactors. Below we summarize the main findings of this work.
\begin{itemize}
    \item Ice chemistry in rate equation based astrochemical models for prestellar objects is mainly sensitive to diffusion of small radicals such as \ch{H}, \ch{N}, \ch{O}, \ch{CH3}, and \ch{HCO}. More attention needs to be given to accurate calculation or measurement of the diffusion barrier of these larger radicals.
    \item In many cases, accurate energy barriers for reactions are not essential, because diffusion is actually the rate-limiting step. This means that reaction energy barriers can be determined using more approximate methods. Depending on whether it is faster or not than the hopping rate, a more accurate barrier needs to be determined.
    \item Fitting binding energies, diffusion barriers or reaction energy barriers from observations is extremely risky because of the complex interplay of many parameters. When trying to constrain parameters, the observed species need to be sensitive to the parameters.
    \item Care needs to be taken when fitting observed abundances to obtain physical conditions because the modeled abundances of some species are highly sensitive to the chemical parameters in the network. This could lead to differences in determined physical conditions when using different astrochemical codes or networks.
\end{itemize}

\begin{acknowledgements}
T.M.D. is financially supported by the Dutch Astrochemistry Network of the Dutch Research Council (NWO) under grant no.~ASTRO.JWST.001. S.V. acknowledges funding from the European Research Council (ERC) under the European Union’s Horizon 2020 research and innovation programme MOPPEX 833460. The authors thank Kenji Furuya and Germ\'an Molpeceres for fruitful discussions.
\\
\textit{Software.}~UCLCHEM \citep{Holdship2017}, NumPy \citep{Harris2020}, Matplotlib \citep{Hunter2007}, SciPy \citep{scipy}, pandas \citep{pandas2020}.
\end{acknowledgements}

\bibliographystyle{aa}
\bibliography{references}

\begin{appendix}
\section{Estimation of tunneling mass}
\label{app:tunneling_mass}
For a Langmuir-Hinshelwood reaction, the mass of the tunneling ``particle'' is essential in correctly calculating the rate constant. In most astrochemical models, the mass of the tunneling particle is taken as the reduced mass of the two reagents \citep{Hasegawa1992,Ruaud2016},
\begin{equation}
    \mu^{i+j} = \frac{m^i m^j}{m^i+m^j}.
    \label{eq:naiveReducedMass}
\end{equation}
However, this strongly overestimates the mass of the tunneling particle (and so underestimates the rate constant) for some reactions. For example, the reagents of reaction \ch{CH4 + OH -> CH3 + H2O} have a reduced mass of $\sim$8~amu, while the tunneling particle is a hydrogen atom (moving from \ch{CH4} to the \ch{OH} radical) and so to get an accurate tunneling probability a tunneling particle mass of 1~amu should be used instead. With a barrier of 2575~K \citep{Lamberts2017}, using the reduced mass results in a tunneling probability underestimated by $\sim$23 orders of magnitude, whereas using $\mu=1$~amu underestimates the reaction rate constant by about 4 orders of magnitude. To remedy this, we estimate the tunneling mass of a Langmuir-Hinshelwood reaction using the following method, similar to \citet{Garrod2017}. For a one-product reaction, we simply use Eq.~\ref{eq:naiveReducedMass}. For a two-product reaction, find which reagent is most similar to which product. If the difference between a reagent and the corresponding product is one atom, that atom has been exchanged between the two reagents, and so the tunneling mass is set to the mass of that atom. If not, Eq.~\ref{eq:naiveReducedMass} is used. This method does not always decrease the tunneling mass. For example, for \ch{H + O2H -> 2 OH}, the predicted tunneling mass is $m^{\text{O}}=16$~amu, whereas the reduced mass of the reagents is $\sim$1~amu. Barrierless reactions are unaffected because their tunneling probability is unity, regardless of the tunneling mass.

\FloatBarrier

\section{Enthalpies of formation}
\label{app:formation}
\begin{figure}
  \capstart
  \centering
  \resizebox{6.8cm}{!}{\includegraphics{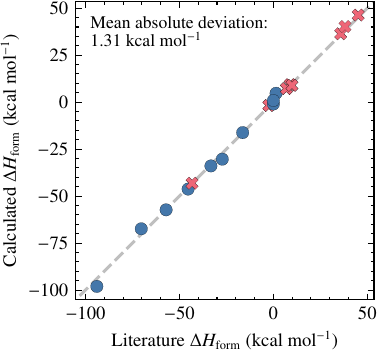}}
  \caption{Comparison of enthalpies of formation from ATcT and calculated using the method described in App.~\ref{app:formation}. Blue circles indicate molecules with an even number of electrons, whereas red crosses indicate molecules with an odd number of electrons. The dashed gray line indicates a perfect agreement.}
  \label{appfig:enthalpies_accuracy}
\end{figure}

If the enthalpy of formation of a species was unavailable in databases, it was instead calculated using quantum chemical methods. First, the (approximately) optimal geometry was generated from the SMILES code using OpenBabel's 3D conformer generation \citep{OBoyle2011,Yoshikawa2019}. The precision of the generated atomic coordinates was then reduced to 3 decimal places to prevent ORCA from placing incorrect symmetry constraints during further geometry optimizations. High accuracy energies were then calculated for the molecules and the atoms they are composed of using the G2(MP2,SVP) method, as described by \citet{Curtiss1996}. This method consists of geometry optimizations, a zero-point vibrational energy calculation, and high-accuracy single-point calculations, resulting in $E_0$. The calculations were carried out in ORCA version 6.1.0 \citep{Neese2025}, with the \texttt{compound[G2-MP2-SVP]} keyword. This is done for the two lowest multiplicities (i.e., multiplicities of 1 and 3 for a species with an even number of electrons, and multiplicities of 2 and 4 for species with an odd number of electrons) to find the true electronic ground state. The enthalpy of formation of a molecule $\mathrm{A}_x\mathrm{B}_y\mathrm{C}_z$ is then
\begin{align}
    \Delta H_{\mathrm{form}}^{\mathrm{A}_x\mathrm{B}_y\mathrm{C}_z}=E_0^{\mathrm{A}_x\mathrm{B}_y\mathrm{C}_z} &+ x\left(\Delta H_{\mathrm{form}}^{\mathrm{A}}-E_0^{\mathrm{A}}\right) \nonumber \\
    &+ y\left(\Delta H_{\mathrm{form}}^{\mathrm{B}}-E_0^{\mathrm{B}}\right) + z\left(\Delta H_{\mathrm{form}}^{\mathrm{C}}-E_0^{\mathrm{C}}\right),
\end{align}
where the enthalpies of formation of atoms A, B and C are experimental values taken from the Active Thermochemical Tables \citep[ATcT,][]{ATcT}. The effect of a non-zero temperature is neglected. Figure~\ref{appfig:enthalpies_accuracy} compares literature values for the enthalpies of formation from ATcT with values calculated with G2(MP2,SVP). This method is clearly accurate enough, even for open-shell molecules, with a mean absolute deviation of only 1.31~kcal~mol$^{-1}$, which is just above ``chemical accuracy''. The enthalpies of formation are only used here for determining the chemical desorption probability, and are therefore less critical.
The newly calculated enthalpies of formation are shown in Table~\ref{tab:enthalpies_of_formation}. The code to calculate enthalpies of formation of new species is available on GitHub\footnote{\url{https://github.com/TobiasDijkhuis/HEAT_calculator}}.
\begin{table}
\caption{Newly calculated enthalpies of formation using the method described in App.~\ref{app:formation}.}
\label{tab:enthalpies_of_formation}
\centering
\begin{tabular}{l l c S[table-number-alignment = center]}
    \hline\hline
Species & SMILES & Multiplicity & \multicolumn{1}{c}{$\Delta H_{\mathrm{form}}$ (kcal~mol$^{-1}$)} \\
\hline
    \ch{HCSH} & [H]=[C]S & 1 & 72.60 \\
    \ch{NSH} & [N][S][H] & 1 & 74.47 \\
    \ch{H2NS} & N[S] & 2 & 35.1 \\
    \ch{HNSH} & [H][N]S & 2 & 48.90 \\
    \ch{NSH2} & [N]=[SH2] & 2 & 98.93 \\
    \ch{CH2SH2} & C=[SH2] & 1 & 59.62 \\
    \ch{HSO} & S[O] & 2 & -4.18 \\
    \ch{HNSH2} & N=[SH2] & 1 & 56.20 \\
    \ch{OCSH} & O=[C]S & 2 & 5.77 \\
    \ch{OCHS} & O=C[S] & 2 & 6.47 \\
    \ch{HOCS} & O[C]=S & 2 & 18.80 \\
    \ch{HSO2} & O=S[O] & 2 & -30.81 \\
    \ch{HOSO} & O[S]=O & 2 & -54.64 \\
    \ch{HS2} & S\#[S] & 2 & 27.50 \\
    \hline
\end{tabular}
\end{table}
\FloatBarrier

\section{Complex organic molecules}
\label{app:complexOrganicMolecules}
Complex organic molecules (COMs) are organic molecules consisting of 6 or more atoms. They are believed to be important precursors for the formation of biologically relevant molecules that might lead to the emergence of life.
\begin{figure*}
    \centering
    \capstart
   \includegraphics[width=18cm]{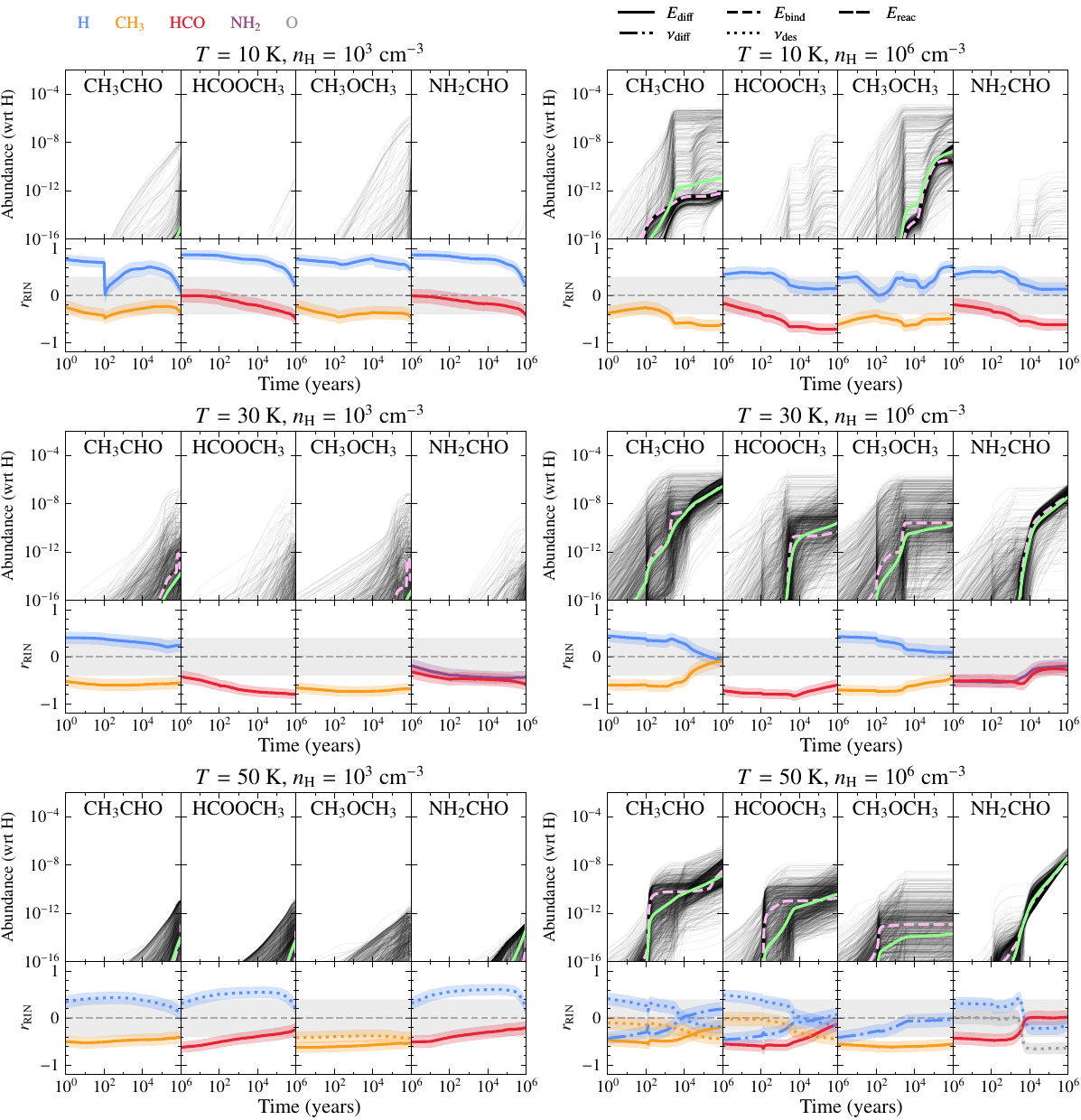}
     \caption{Same as Fig.~\ref{fig:VariousPhysicalConds} but for various complex organic molecules.}
     \label{fig:VariousPhysicalCondsCOMs}
\end{figure*}

Figure~\ref{fig:VariousPhysicalCondsCOMs} shows sensitivities at different physical conditions for various COMs. At low temperature (10~K), many COMs are sensitive to the diffusion barrier of atomic H. If the H diffusion barrier is higher, their abundance is too. These COMs are mostly formed by barrierless recombination of smaller radicals, such as \ch{CH3 + CHO -> CH3CHO}. The abundance of these radicals is governed by the rate of their hydrogenation, for example \ch{CH3 + H -> CH4} or \ch{CHO + H -> H2CO / CO + H2}. Because these larger radicals are more strongly bound to the surface than atomic hydrogen, the reaction rate is mainly determined by the \ch{H} diffusion rate. If the hydrogen diffusion barrier is higher, these hydrogenation reactions will occur slower and as a result the abundance of these radicals are larger, such that the abundance of the resulting COMs is also larger. This was also found by \citet{Furuya2024} and \citet{Jin2020}.

At 30~K, the abundance of \ch{H} on the ice is low and its diffusion occurs rapidly, and the limiting step becomes the diffusion of the larger radicals. This means that many COMs are sensitive to the diffusion barriers of the reagent radicals. For example, \ch{CH3CHO} is very sensitive to the \ch{CH3} diffusion barrier, because it is the more mobile one of the two reagents that react to form it. On the other hand, \ch{HCOOCH3} is more sensitive to the diffusion barrier of \ch{HCO} than the one of \ch{CH3O} because \ch{CH3O} is more strongly bound to the surface.

At 50~K, the situation is similar, but now also the desorption of atomic hydrogen is important. Higher desorption prefactors lead to a lower abundance of atomic hydrogen on the ice, which leads to a lower rate of radical destruction. This in turn, as was the case at 10~K, leads to a higher abundance of COMs.

\begin{figure}
  \capstart
  \resizebox{\hsize}{!}{\includegraphics{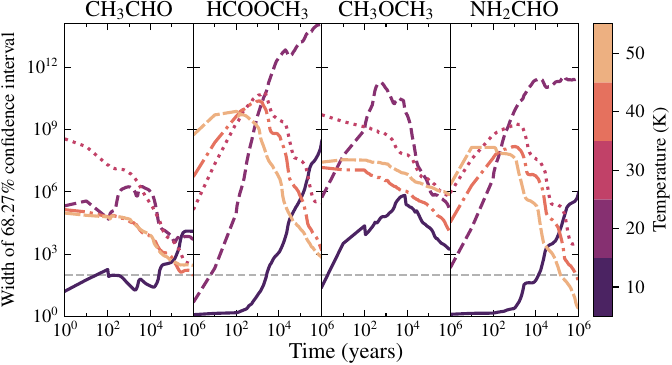}}
  \caption{Width of the 68.27\% confidence intervals of abundances of some complex organic molecules at various temperatures (averaged over all densities) over time at $\zeta=1.3\times10^{-17}$~s$^{-1}$ and $F_{\mathrm{UV}}=1$~Habing. The solid, dashed, dotted, dashed-dotted, and densely dotted lines correspond to 10, 20, 30, 40 and 50~K, respectively. The dashed gray line indicates a width of 2 orders of magnitude.}
  \label{fig:confidence_intervals_abundances_COMs}
\end{figure}

Figure~\ref{fig:confidence_intervals_abundances_COMs} shows the widths of the 68.27\% confidence intervals of abundances. This shows that these complex organic molecules are extremely sensitive to the network being used. Especially at 20~K, where diffusion of \ch{HCO} is either possible or not possible depending on its diffusion barrier, the uncertainties of \ch{HCOOCH3} and \ch{NH2CHO} are extremely large.

\FloatBarrier

\section{Variation of reaction energy barriers}
\label{app:variation_of_reaction_energy_barriers}
To confirm that the abundances are not sensitive to the reaction energy barriers, we repeated the same calculations, but only varying the reaction energy barriers and keeping the binding energies, diffusion barriers and the prefactors fixed to their nominal values. Additionally, we widen the spread of energy barriers to enlarge the range of sampled rate constants $\kappa^{i+j}$, as shown in Table~\ref{tab:uncertaintiesParametersOnlyReactions}. These calculations were done on a grid of only varying the temperature and density because the cosmic ray ionization rate and UV field strength have a negligible effect on the correlations, as discussed in Sect.~\ref{subsec:across_physical_conditions}.
\begin{table}
\caption{Sample widths for reaction energy barriers..}
\label{tab:uncertaintiesParametersOnlyReactions}
\centering
\begin{tabular}{
    c 
    S[table-format = <4,table-text-alignment=left]@{~}
    l
    S[table-format = 4, table-text-alignment=right]@{~}
    l
    }
\hline\hline
Parameter type & \multicolumn{2}{c}{$\mu$} & \multicolumn{2}{c}{$\sigma$} \\
\hline
    Reaction barrier & {$< 200$} & K & {100} & K \\
           & {$\leq 1600$} & K & \multicolumn{2}{c}{$\frac{4}{5}\mu$} \\
           & {$> 1600$} & K & {1280} & K \\
\hline
\end{tabular}
\tablefoot{Energies are at least 0~K. All other parameters are kept fixed.}
\end{table}

\begin{figure}
  \capstart
  \resizebox{\hsize}{!}{\includegraphics{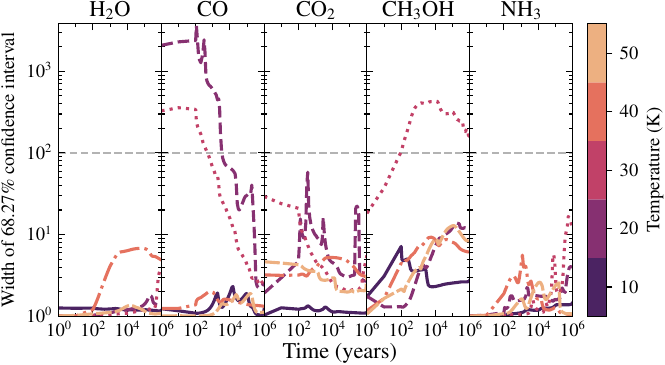}}
  \caption{Width of the 68.27\% confidence intervals of abundances of selected ice species while only varying the reaction energy barriers at various temperatures (averaged over all densities) over time at $\zeta=1.3\times10^{-17}$~s$^{-1}$ and $F_{\mathrm{UV}}=1$~Habing. The solid, dashed, dotted, dashed-dotted, and densely dotted lines correspond to 10, 20, 30, 40 and 50~K, respectively. The dashed gray line indicates a width of 2 orders of magnitude.}
  \label{fig:confidence_intervals_abundances_reactionsOnly}
\end{figure}

Figure~\ref{fig:confidence_intervals_abundances_reactionsOnly} shows the widths of the 68.27\% confidence intervals of the main ice species, varying only the reaction energy barriers. The confidence intervals of \ch{H2O}, \ch{CO2} and \ch{NH3} are quite narrow, indicating that the abundances are unaffected by changing the reaction energy barriers. The abundances of \ch{CO} and \ch{CH3OH} at 20 and 30~K still have a spread of about 3 orders of magnitude at most. This is mostly because of variation of the energy barrier of \ch{H + CO -> HCO}, with $r_{\mathrm{RIN}}\approx0.8$ at 30~K and $10^6$~cm$^{-3}$ from $10^3$ to $10^6$~years. Nevertheless, it can be seen that the variance in abundances is much smaller, often within 1~order~of~magnitude.

\end{appendix}
\end{document}